\documentclass[apjl]{emulateapj}

\def\Lsun{\hbox{\it L$_\odot$}}

\def\Msun{\hbox{\it M$_\odot$}}
\def\Minit{\hbox{\it M$_{\rm initial}$}}

\def\Myr{\hbox{\it Myr}}

\def\kms{\hbox{km$\,$s$^{-1}$}}

\def\simgr{\mathrel{\hbox{\rlap{\hbox{\lower4pt\hbox{$\sim$}}}\hbox{$>$}}}}

\def\lbv{\hbox{LBV~1806$-$20}}
\def\sgr{\hbox{SGR~1806$-$20}}

\shorttitle{SGR 1806$-$20 CLUSTER}
\shortauthors{FIGER ET AL.}

\begin{document}

\title{MASSIVE STARS IN THE \sgr\ CLUSTER}

\author{
Donald F. Figer\altaffilmark{1}, 
Francisco Najarro\altaffilmark{2},
T. R. Geballe\altaffilmark{3}, 
R. D. Blum\altaffilmark{4},
Rolf P. Kudritzki\altaffilmark{5}
}

\email{figer@stsci.edu}

\altaffiltext{1}{Space Telescope Science Institute, 3700 San Martin Drive, Baltimore, MD 21218; figer@stsci.edu }
\altaffiltext{2}{Instituto de Estructura de la Materia, CSIC, Serrano 121, 29006 Madrid, Spain }
\altaffiltext{3}{Gemini Observatory, 670 N. A'ohoku Pl., Hilo, HI 96720}
\altaffiltext{4}{Cerro Tololo Interamerican Observatory, Casilla 603, La Serena, Chile}
\altaffiltext{5}{Institute for Astronomy, University of Hawaii, 2680 Woodlawn Drive, Honolulu, HI 96822}

\begin{abstract}
We report the discovery of additional hot and massive stars in the cluster surrounding the soft
gamma repeater
SGR~1806$-$20, based upon UKIRT and Keck near-infrared spectroscopy. Of the newly
identified stars, three are Wolf-Rayet stars of types WC8, WN6, and WN7, and a fourth
star is an OB supergiant.  These three stars, along with four previously discovered,
imply a cluster age of $\sim$3.0-4.5~\Myr, based on the presence of WC stars and
the absence of red supergiants. Assuming coevality, this age suggests that the
progenitor of \sgr\ had an initial mass greater than $\sim$50~\Msun. 
This is consistent with the suggestion that SGRs are post-supernovae end states of massive
progenitors, and may suggest that only massive stars evolve into magnetars that
produce SGRs. It also suggests that very massive stars can evolve into neutron stars,
not just black holes, as recently predicted by theory. The cluster age also provides
constraints on the very high mass object, \lbv.  

\end{abstract}
\keywords{stars: evolution --- stars: neutron --- stars: individual (LBV1806-20) --- stars: individual (SGR 1806$-$20) --- infrared: stars}

\section{Introduction}
Massive stellar clusters are the birth sites of the most massive stars, and are
proving grounds for theories of massive star formation, evolution, end states, 
and Galactic chemical enrichment. With the recent advent of
sensitive infrared instrumentation, on ground and space based platforms, many massive
clusters in the Galaxy are now just being discovered and investigated. At least two of
these clusters contain soft gamma repeaters (SGRs), a rare phenomenon characterized
by persistent and energetic bursts of gamma ray emission lasting up to several seconds
and releasing $\sim$10$^{40}$~ergs~s$^{-1}$ \citep{maz79a,maz79b,kou98}. Only four
SGRs are known, three in the Galaxy and one in the SMC. Three
of the SGRs are associated with massive stellar clusters.  

SGR~1806$-$20 is surrounded by a cluster of massive stars \citep{fuc99}.  
\citet{kul95} identified a luminous star that they claimed may be the infrared
counterpart to \sgr, but \citet{eik01} and \citet{kap02} determined that it is too
far from the Chandra error box; no near-infrared counterpart has yet been found, even
down to very faint magnitudes \citep{kou04,isr04}. \citet{van95} find that
this star has characteristics
similar to those of Luminous Blue Variables (LBVs), i.e.\ it has L$\simgr$10$^6$~\Lsun\ and a
spectral type in the range O9-B2. \citet{eik04} estimated an extremely high mass for
\lbv, $\sim$200~\Msun; but \citet{fig04} have shown that it has double lines and
thus may be binary.

In this Letter we present near-infrared spectra of fifteen bright sources near \sgr,
We find that three of them are massive stars in the advanced Wolf-Rayet stage of
post-main sequence evolution, and another three are less evolved OB supergiant stars.
We use these results to constrain the nature of \sgr.

\section{Observations and Data Reduction}

Data were obtained at UKIRT\footnote{The United Kingdom Infrared Telescope (UKIRT) is
operated by the Joint Astronomy Centre on behalf of the Particle Physics and Astronomy
Research Council.} on 7 and 8 June 2004, using UIST \citep{ram98} in medium resolution
mode (R$\sim$800=$\lambda/\Delta\lambda_{\rm FWHM}$, 0$\farcs$24 slit width)  from
1.50~\micron\ to 2.50~\micron. One or two 60 second exposures were obtained,
after which the telescope was nodded along the direction of the slit and the similar
exposures obtained. Star HIP86814 (F6V) was observed as a telluric standard; its
Brackett series lines were removed by interpolation prior to ratioing.

\begin{figure}
\epsscale{1.4}
\plotone{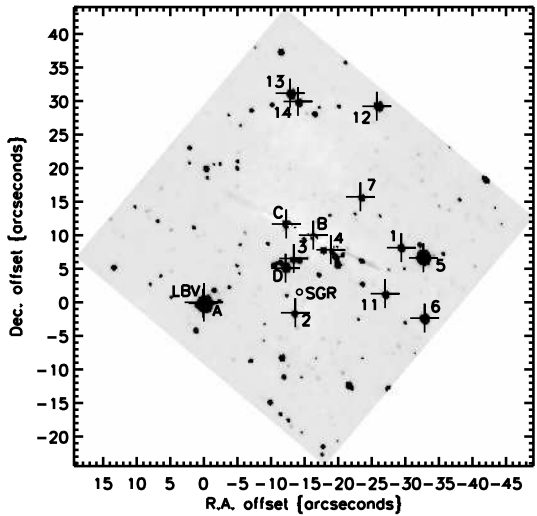}
\caption
{\label{fig:cont} NIRSPEC/SCAM image of stars in the 1806$-$20 cluster with coordinates
offset from \lbv\ at 18$^h$08$^m$40.312$^s$ $-$20$\arcdeg$24$\arcmin$41$\farcs$14 (J2000).
Number designations are from this paper. Letter designations are from \citet{eik04}.
Stars \#8, 9, and 10 are to the west of the field. The slit
can be seen as a linear artifact superpositioned on images of Star ``C'' and ``B.'' A
circle marks the location of \sgr \citep{kap02}.}
\end{figure}

Additional imaging and spectroscopic data were obtained at Keck\footnote{Data
presented herein were obtained at the W.M. Keck Observatory, which is operated as a
scientific partnership among the California Institute of Technology, the University of
California and the National Aeronautics and Space Administration.  The Observatory was
made possible by the generous financial support of the W.M. Keck Foundation.} on 22
June 2003, using NIRSPEC \citep{mcl98}, with the NIRSPEC-6 filter, which covers the H
and K bands). The slit-viewing camera (SCAM) exposures were 2 seconds and provided
stellar images having FWHM$\sim$0$\farcs$4.  Figure~\ref{fig:cont} shows a SCAM image;
note the faint image produced by the slit.  Spectra covering 1.6--2.35~\micron were
obtained at high resolution (R$\sim$26,500=$\lambda/\Delta\lambda_{\rm FWHM}$,
0$\farcs$45 slit width), using two 100 second exposures per grating setting,
with the telescope nodded along the direction of the slit length in between exposures.  
Star \#3 of the Quintuplet cluster was observed as a telluric standard \citep{fig99}.

For both sets of spectra arc lamps were observed to set the wavelength scale and
continuum lamps provided ``flat'' images. Data were reduced by differencing the nod  
pair frames to remove background flux, dark current, and residual bias 
and dividing the result by the flat field frame in order to compensate
for non-uniform response and illumination of the array. Finally, each extracted 
target spectrum was divided by the spectrum of a telluric standard.

Table~1 lists the coordinates, magnitudes, spectral types, and dates observed for the
massive stars and field red giants. 
Figures~\ref{fig:wrspectra}, \ref{fig:hei}, and \ref{fig:otherspectra}
show spectra of these stars.

\begin{deluxetable}{rrrrrrl}
\tablewidth{0pt}
\tablecaption{Massive Stars in the 1806$-$20 Cluster}
\tablehead{
\colhead{ID} & 
\colhead{R.A.} & 
\colhead{Dec.} & 
\colhead{H$-$K} & 
\colhead{K} & 
\colhead{Type} & 
\colhead{Dates Observed} }
\startdata
1 & 38.32 & 33.5  & NA & 11.76 & WC8 & 7 June 2004 \\
2 & 39.42 & 42.57  &1.87    & 12.16    & WN6 & 8 June 2004 \\
3 & 39.50 & 35.88  & NA & 12.87 & WN7? & 8 June 2004 \\
4 & 39.16 & 32.88  & NA & 12.45 & OBI & 22 June 2003 \\
5	&	38.13	&	34.77	&	1.68	&	9.25	& RG & 7 June 2004\\
6	&	38.12	&	43.35	&	1.42	&	10.96	& RG & 7 June 2004\\
7 & 38.75 & 26.48  & 1.49 & 11.90 & OBI? & 7 June 2004 \\
8	&	36.72	&	54.27	&	1.40	&	8.70		& RG & 8 June 2004 \\
9	&	36.72	&	27.80	&	1.45	&	8.89		& RG & 8 June 2004 \\
10	&	37.96	&	16.47	&	2.95	&	10.09		& RG & 8 June 2004 \\
11 & 38.52 & 39.95  & 1.66 & 11.92 & OBI? & 8 June 2004 \\
12	&	38.57	&	13.55	&	1.42	&	10.81		& RG &	8 June 2004 \\
13	&	39.49		&	11.77&	NA	&	11.16		& NA &	8 June 2004 \\
14	&	39.40		&	13.02	&	NA	&	12.00		& RG &	8 June 2004 \\
A & 40.31 & 41.14  & NA & 9.26 & LBV & 22 June 2003 \\
B & 39.24 & 31.86  & 2.94 & 10.50 & WC9 & 22 June 2003 \\
C & 39.51 & 30.02  & 1.85 & 10.96 & OBI & 22 June 2003 \\
D & 39.51 & 35.91  & 1.73 & 11.11 & OBI & 22 June 2003 \\
\enddata
\tablecomments{Number designations are from this paper. Letter designations are from \citet{eik04}.
Coordinates are offset from Right Ascension 18$^h$08$^m$ and
Declination $-$20$^{\arcdeg}$24$^{\arcmin}$ (J2000), and are from the 
2MASS PSC, except for \#3, \#4, \#13, and \#14 which were estimated from the Keck SCAM image. Photometry are from
the PSC, except for \#1, \#3, \#4, \#13, \#14, and ``A'' which are estimated from the Keck SCAM image, and
``B'' and ``D'' which are from \citet{eik04}.
The spectral type for ``B'' is taken from \citet{eik04}.
Star \#2 is source ``E'' in \citet{eik01}. The ``RG'' label refers to red giants.}
\end{deluxetable}

\section{Discussion}

\subsection{Photometry and Spectral Analysis}

Photometry was mostly obtained from the 2MASS \footnote{This publication makes use
of data products from the Two Micron All Sky Survey, which is a joint project of the
University of Massachusetts and the Infrared Processing and Analysis Center/California
Institute of Technology, funded by the National Aeronautics and Space Administration
and the National Science Foundation.} \citep{skr97} Point Source Catalog (PSC). Stars
\#1, \#3, and \#4 are not in the catalog, so we estimated photometry for them from the
Keck SCAM image using the 2MASS PSC to set a zero point. As a check, photometry for
stars \#7 and \#11 from the Keck SCAM image differ from the PSC by +0.11 and $-$0.07
magnitudes, respectively. Note that our method relies on the fact that the NIRSPEC-6
filter transmission function, convolved with the spectral energy distributions with
the appropriate extinction of the target stars, gives similar photometry as measured
with the K filter used in 2MASS. Star ``B'' and ``D'' are in crowded regions, so we
use photometry for them from \citet{eik04}, instead of the PSC. As a further
consistency check on the photometric systems used in the Table, note that 2MASS and
\citet{eik01} photometry match to within the quoted uncertainties of 0.1 magnitudes
for Star \#2.

We estimated spectral types by comparing the observed spectra to those
in the Wolf-Rayet spectral atlas of \citet{fig97} and \citet{han96}.

\begin{figure}
\epsscale{1.01}
\plotone{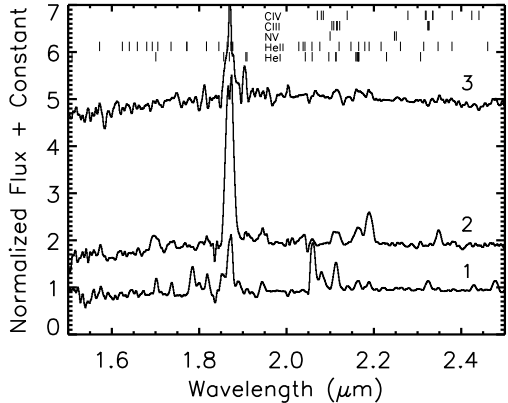}
\caption
{\label{fig:wrspectra} UKIRT/UIST spectra of the newly discovered WR stars. The number
labels refer to objects in Table~1.}
\end{figure}

The spectra in Figure~\ref{fig:wrspectra} reveal three new WR stars.  Star \#1 has
strong lines of \ion{C}{3} (2.11~\micron), \ion{C}{4} (2.08~\micron), \ion{He}{1}
(1.70~\micron\ and 2.06~\micron), and \ion{He}{2} (1.87~\micron\ and 2.16~\micron).  
Taken together, these lines indicate a classification of WC8. Stars \#2 and \#3 have
strong lines of \ion{He}{2}; neither has appreciable \ion{N}{5} (2.10~\micron)
emission. Star \#2 has relatively strong \ion{He}{2} (1.87~\micron\ and
2.16~\micron), and EW$_{2.189~\micron}$/EW$_{2.112~\micron}$=2.3, consistent with
a WN6 spectral type \citep{fig97}.
For star \#3, we
suggest a classification of WN7, but note that the spectrum has relatively low S/N.

\begin{figure}
\epsscale{1.6}
\plottwo{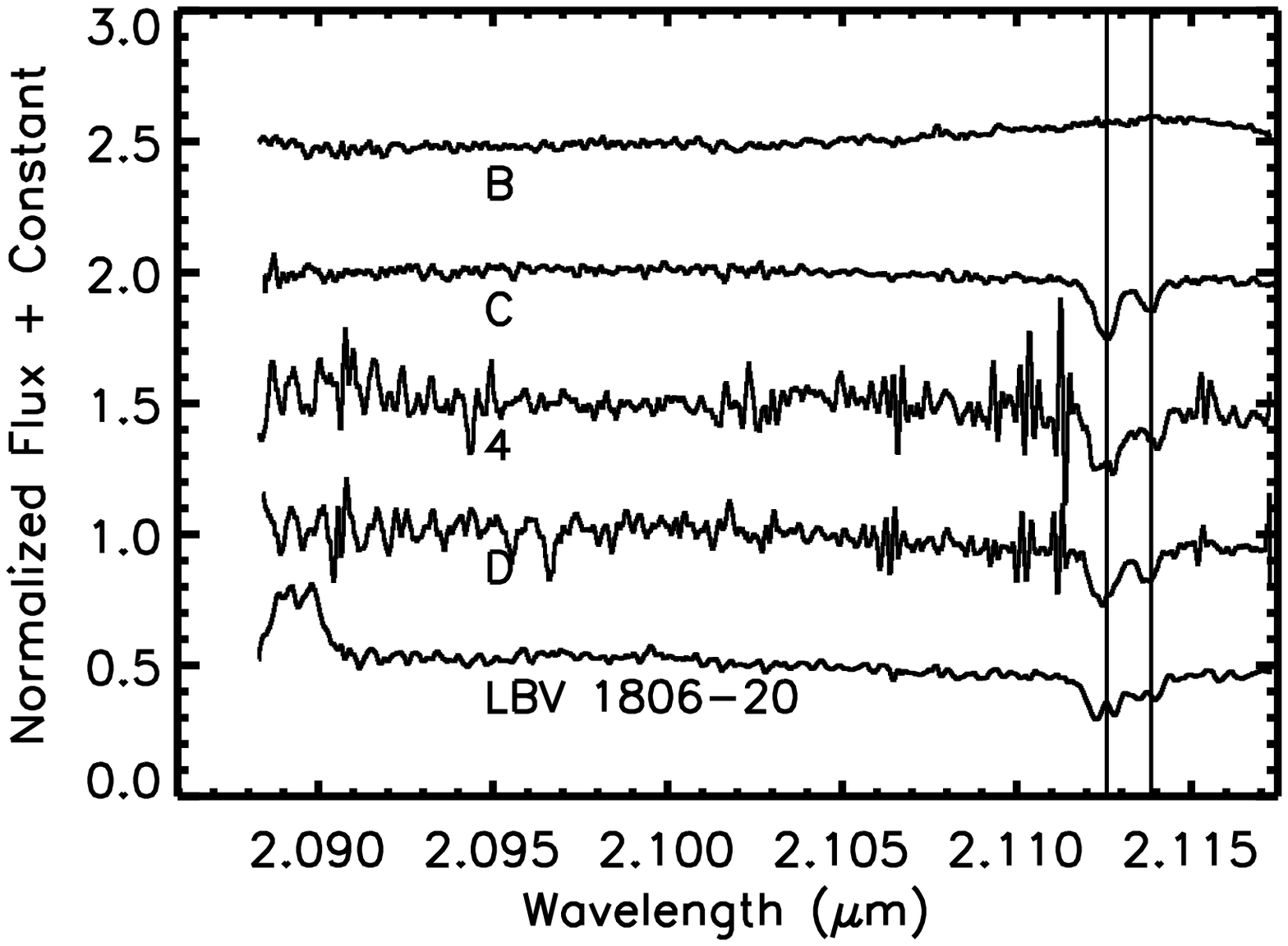}{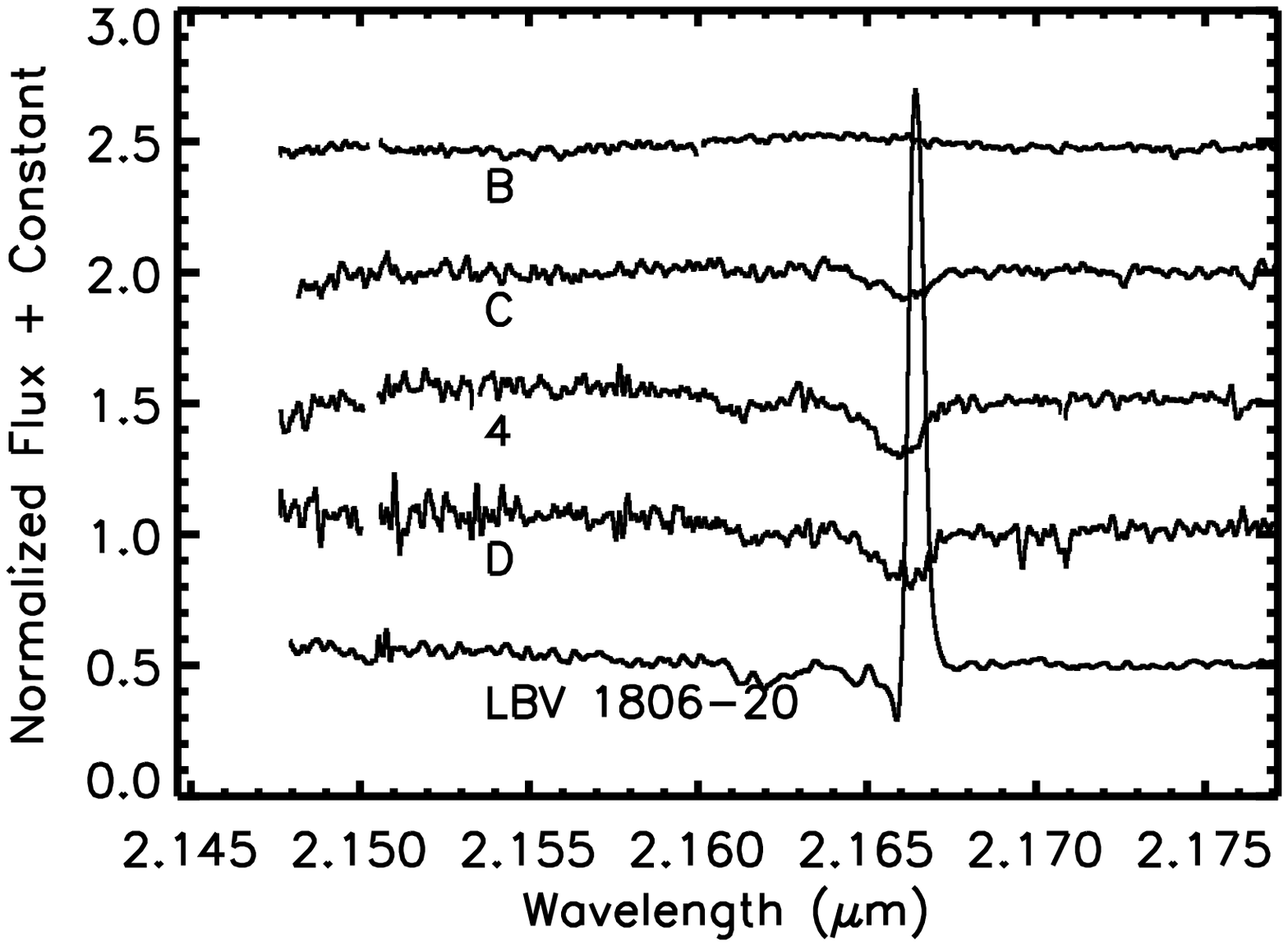}
\caption
{\label{fig:hei} Keck/NIRSPEC spectra of massive stars near the 2.112/2.113~\micron\  \ion{He}{1} doublet ({\it left}),
and the 2.166~\micron\ Br-$\gamma$ line ({\it right}). 
The spectra (except for ``B'') show the 
\ion{He}{1} doublet in absorption near 2.1125~\micron\ is marked 
by vertical lines. This feature in \lbv\ is obviously double.
The broad feature near 2.089~\micron\ is common to LBV spectra and corresponds to an \ion{Fe}{2} transition.
Weak \ion{He}{1} absorption lines just short of 2.166~\micron\ are visible in all spectra, except for ``B.''}
\end{figure}

The spectra in Figure~\ref{fig:hei} show three less evolved OBI stars (``4'', ``C'',
and ``D''), as well as the evolved luminous blue variable, \lbv, and a previously
identified WR star, ``B'' \citep{eik04}. \lbv\ is discussed at length in
\citet{fig04}, and references therein. The OBI stars show modest \ion{He}{1} and
\ion{H}{1} absorption components with equivalent widths of
EW$_{2.112~\micron}$$\sim$2-3~\AA\ and EW$_{2.166~\micron}$$\sim$3-5~\AA, and
FWHM$\sim$70-80~\kms; both line depths and equivalent widths are consistent with the
classifications of O9-B3 for these stars \citep{han96}. Note that our high-resolution
spectra successfully resolve these individual lines that would otherwise
appear as blends in lower resolution spectra. Indeed, our 
linewidths are significantly less than those estimated in much lower spectral
resolution data of \citet{eik04}, and they do not support
the suggestion that these stars are extraordinarily luminous hypergiants. 

\begin{figure}
\epsscale{1.1}
\plotone{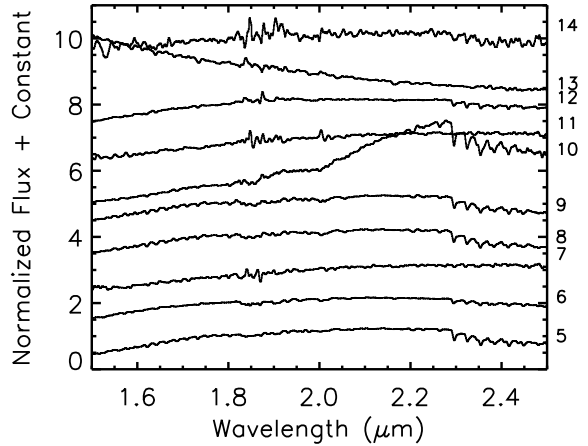}
\caption
{\label{fig:otherspectra} UKIRT/UIST spectra of stars within and near the field
in Figure~\ref{fig:cont}. Most of the spectra have CO bandhead absorption starting
at 2.293~\micron\ and characteristic of red giants.}
\end{figure}

Figure~\ref{fig:otherspectra} demonstrates that most of the remaining sources are red
giants (RG's), given their strong CO absorptions at 2.3~\micron\ and beyond. These
stars are relatively old ($\tau_{\rm age}>$100~Myr) and have no physical association
with the cluster. Stars \#7
and \#11 lack CO and thus could be OBI~stars, but they lack Br-$\gamma$ absorption
lines. Pa-$\alpha$ absorption lines in their spectra would be comparable to the noise
levels, so it is difficult to rule out their existence.

\subsection{Massive Stars and \sgr}

The 1806$-$20 cluster contains at least four WR stars, three OBI stars, and an
LBV-candidate. While such a collection of spectral types may seem extraordinary, it is
expected from stellar evolution models.  
The non-rotating Geneva models predict the onset of red supergiants 
at $\sim$4-4.5~\Myr\ and WC stars at $\sim$3.0-3.5~\Myr\ for solar to twice-solar 
metallicity \citep{mey94}. Non-zero rotation rates tend to increase both limits 
by $\sim$20\% \citep{mey05}. Given the presence of WC stars and absence of red supergiants, we estimate a cluster age of
$\sim$3.0-4.5~\Myr, assuming coeval formation. 
Appealing to the Geneva models, we should then expect to still be able
to see stars with initial masses $\sim$50-120~\Msun\ in the cluster. 
Our age estimate indicates a somewhat lower mass for \lbv\ than given by \citet{eik04} since 
such massive progenitors typically have ages less than 3~\Myr\ \citep{bon84}. A lower 
mass is also consistent with the claim by \citet{fig04} that \lbv\ may 
be a binary system having \Minit$\sim$130~\Msun.

The 1806$-$20 cluster is similar in spectral content to the Quintuplet cluster
\citep{fig99}, albeit at a much smaller mass. Both have roughly equal numbers of WC
and WN stars, although the latter may have twice as many WC's if the five very red
sources for which the Quintuplet was named are indeed dusty WC stars. One minor
difference is the presence of one red supergiant in the Quintuplet and none in the
1806$-$20 cluster. This may be the consequence of low number statistics, as the
Quintuplet cluster has four times more massive stars; or it may be an indication that
the 1806$-$20 cluster is slightly younger than the 4~\Myr\ age estimated for the
Quintuplet. Assuming 8 stars with \Minit$>$20~\Msun, we estimate M$_{\rm
cluster}>$1,750~\Msun, assuming a Salpeter IMF down to 0.8~\Msun\ \citep{sal55}. If WR
and LBV stars have \Minit$>$50~\Msun, then M$_{\rm cluster}>$6,600~\Msun.  The
supernova producing \sgr\ likely could not have triggered formation of the cluster
\citep{eik01}, because the magnetar is thought to be $<$10$^4$ yrs old, while the
cluster stars are $\sim$4~\Myr\ old.  

A further important conclusion is that the masses and ages of the cluster stars are
consistent with the hypothesis that \sgr\ represents the end state of a particularly
massive cluster member. Because more massive stars reach their end states more quickly
than less massive stars, the likely mass of the SGR progenitor must be greater than
presently observed masses. Assuming that the presently observed stars had
\Minit$\gtrsim$50~\Msun, \sgr\ had a larger initial mass, as suggested by
\citet{kap02}.  SGR~1900+14 and SGR~0526$-$66 are also associated with massive stellar
clusters \citep{vrb00,klo04}, suggesting that magnetars may be short-lived
descendants of massive stars.  These three cases support the theoretical prediction
that very massive stars lose most of their mass and collapse as neutron stars, not
black holes \citep{fry01,woo02}. They may also support a model in which magnetars are
preferentially produced by stars with large initial mass \citep{eik04}.

\acknowledgements
We acknowledge useful discussions with Jeff Valenti, Andy Fruchter, and Stephan Eikenberry.
We thank Keck staff members Randy Campbell and Grant Hill for assistance with
the observations. F.~N. acknowledges grants AYA2003-02785-E and ESP2002-01627.
Finally, we acknowledge a similar result, by independent means, in Gaensler et al.\ (2005).


\small
\begin{thebibliography}{}
\bibitem[Bond, Arnett, \& Carr(1984)]{bon84} Bond, J.~R., Arnett, W.~D., \& Carr, B.~J.\ 1984, \apj, 280, 825 
\bibitem[Eikenberry et al.(2001)]{eik01} Eikenberry, S.~S., Garske, M.~A., Hu, D., Jackson, M.~A., Patel, S.~G., Barry, D.~J., Colonno, M.~R., \& Houck, J.~R.\ 2001, \apjl, 563, L133 
\bibitem[Eikenberry et al.(2004)]{eik04} Eikenberry, S.~S. et al.\ 2004, \apj, 616, 506
\bibitem[Feitzinger, Schlosser, Schmidt-Kaler, \& Winkler(1980)]{fei80} Feitzinger, J.~V., Schlosser, W., Schmidt-Kaler, T., \& Winkler, C.\ 1980, \aap, 84, 50
\bibitem[Figer et al.(2004)]{fig04} Figer, D.~F., Najarro, F., \& Kudritzki, R.~P.\ 2004, ApJ, 610, L109
\bibitem[Figer, McLean, \& Najarro(1997)]{fig97} Figer, D.~F., McLean, I.~S., \& Najarro, F.\ 1997, \apj, 486, 420 
\bibitem[Figer \& Kim(2002)]{fig02} Figer, D.~F.~\& Kim, S.~S.\ 2002, ASP Conf.~Ser.~263: Stellar Collisions, Mergers and their Consequences, 287 
\bibitem[Figer et al.(1999)]{fig99} Figer, D. F., McLean, I. S., \& Morris, M. 1999, \apj, 514, 202 
\bibitem[Figer et al.(1998)]{fig98} Figer, D. F., Najarro, F., Morris, M., McLean, I. S., Geballe, T. R., Ghez, A. M., \& Langer, N. 1998, \apj, 506, 384
\bibitem[Fryer \& Kalogera(2001)]{fry01} Fryer, C.~L.~\& Kalogera, V.\ 2001, \apj, 554, 548 
\bibitem[Fuchs et al.(1999)]{fuc99} Fuchs, Y., Mirabel, F., Chaty, S., Claret, A., Cesarsky, C.~J., \& Cesarsky, D.~A.\ 1999, \aap, 350, 891 
\bibitem[Gaensler et al.(2005)]{gae05} Gaensler, B.~M., McClure-Griffiths, N.~M., Oey, M.~S., Haverkorn, M., Dickey, J.~M., \& Green, A.~J.\ 2005, ArXiv Astrophysics e-prints, astro-ph/0501563 
\bibitem[Geballe et al.(2000)]{geb00} Geballe, T.R., Figer, D.F., \& Najarro, F.\ 2000, \apj, 530, 97
\bibitem[Genzel et al.(1996)]{gen96} Genzel, R., Thatte, N., Krabbe, A., Kroker, H., \& Tacconi-Garman, L. E. 1996, \apj, 472, 153
\bibitem[G{\" o}tz, Mereghetti, Mirabel, \& Hurley(2004)]{got04} G{\" o}tz, D., Mereghetti, S., Mirabel, I.~F., \& Hurley, K.\ 2004, \aap, 417, L45 
\bibitem[Hanson, Conti, \& Rieke(1996)]{han96} Hanson, M.~M., Conti, P.~S., \& Rieke, M.~J.\ 1996, \apjs, 107, 281 
\bibitem[Hurley et al.(1999)]{hur99} Hurley, K., Kouveliotou, C., Cline, T., Mazets, E., Golenetskii, S., Frederiks, D.~D., \& van Paradijs, J.\ 1999, \apjl, 523, L37 
\bibitem[Israel et al.(2004)]{isr04} Israel, G.~L., Mignani, R., Covino, S., Campana, S., Stella, L., Rea, N., Testa, V., \& Marconi, G.\ 2004, GRB Circular Network, 2609, 1 
\bibitem[Kaplan et al.(2002)]{kap02} Kaplan, D.~L., Fox, D.~W., Kulkarni, S.~R., Gotthelf, E.~V., Vasisht, G., \& Frail, D.~A.\ 2002, \apj, 564, 935 
\bibitem[Klose et al.(2004)]{klo04} Klose, S., et al.\ 2004, \apjl, 609, L13 
\bibitem[Koornneef(1983)]{koo83} Koornneef, J.\ 1983, \aap, 128, 84 
\bibitem[Kouveliotou et al.(1998)]{kou98} Kouveliotou, C., Kippen, M., Woods, P., Richardson, G., Connaughton, V., \& McCollough, M.\ 1998, \iaucirc, 6944, 2 
\bibitem[Kouveliotou et al.(2004)]{kou04} Kouveliotou, C., Klose, S., Wachter, S., Woods, P., Patel, S., Greiner, J., Stecklum, B., \& van der Klis, M.\ 2004, GRB Circular Network, 2607, 1 
\bibitem[Kulkarni et al.(1995)]{kul95} Kulkarni, S.~R., Matthews, K., Neugebauer, G., Reid, I.~N., van Kerkwijk, M.~H., \& Vasisht, G.\ 1995, \apjl, 440, L61 
\bibitem[LaVine, Eikenberry, \& Davis(2003)]{lav03} LaVine, J.~L., Eikenberry, S., \& Davis, J.~D.\ 2003, American Astronomical Society Meeting, 203,  
\bibitem[Mazets, Golenetskij, \& Guryan(1979)]{maz79a} Mazets, E.~P., Golenetskij, S.~V., \& Guryan, Y.~A.\ 1979, Soviet Astronomy Letters, 5, 343 
\bibitem[Mazets et al.(1979)]{maz79b} Mazets, E.~P., Golentskii, S.~V., Ilinskii, V.~N., Aptekar, R.~L., \& Guryan, I.~A.\ 1979, \nat, 282, 587 
\bibitem[McLean et al.(1998)]{mcl98} McLean et al.\ 1998, SPIE Vol. 3354, 566
\bibitem[Meynet et al.(1994)]{mey94} Meynet, G., Maeder, A., Schaller, G., Schaerer, D., \& Charbonnel, C.\ 1994, \aaps, 103, 97 
\bibitem[Meynet \& Maeder(2005)]{mey05} Meynet, G., \& Maeder, A.\ 2005, \aap, 429, 581 
\bibitem[Ramsay Howat et al.(1998)]{ram98} Ramsay Howat, S.~K., et al.\ 1998, \procspie, 3354, 456 
\bibitem[Salpeter(1955)]{sal55} Salpeter, E.~E.~1955, \apj, 121, 161
\bibitem[Skrutskie et al.(1997)]{skr97} Skrutskie, M.~F., et al.\ 1997, ASSL Vol.~210: The Impact of Large Scale Near-IR Sky Surveys, 25 
\bibitem[Vacca, Garmany, \& Shull(1996)]{vac96} Vacca, W.~D., Garmany, C.~D., \& Shull, J.~M.\ 1996, \apj, 460, 914 
\bibitem[van Kerkwijk, Kulkarni, Matthews, \& Neugebauer(1995)]{van95} van Kerkwijk, M.~H., Kulkarni, S.~R., Matthews, K., \& Neugebauer, G.\ 1995, \apjl, 444, L33 
\bibitem[Vrba et al.(2000)]{vrb00} Vrba, F.~J., Henden, A.~A., Luginbuhl, C.~B., Guetter, H.~H., Hartmann, D.~H., \& Klose, S.\ 2000, \apjl, 533, L17 
\bibitem[Woosley, Heger, \& Weaver(2002)]{woo02} Woosley, S.~E., Heger, A., \& Weaver, T.~A.\ 2002, Reviews of Modern Physics, 74, 1015 
\end{thebibliography}
\end{document}